# On applications of Ulam-Hyers stability in biology and economics

**E. Ahmed\*, A. M. A. El-Sayed\*\*, H. A. A. El-Saka\*\*\* and Gehan A. Ashry\*\***

\*Mathematics Department, Faculty of sciences,
Mansoura 35516, Egypt.
Email: magd45@yahoo.com

\*\*Mathematics Department, Faculty of Science,
Alexandria University, Alexandria, Egypt
Email: amasayed5@yahoo.com

\*\*\* Mathematics Department, Damietta Faculty of Science,
Mansoura University, 34517, New Damietta, Egypt
Email: halaelsaka@yahoo.com

**Abstract:**

We argue that Ulam-Hyers stability concept is quite significant in realistic problems in numerical analysis, biology and economics. A generalization to nonlinear systems is proposed and applied to the logistic equation (both differential and difference), SIS epidemic model, Cournot model in economics and a reaction diffusion equation. To the best of our knowledge this is the first time Ulam-Hyers stability is considered from the applications point of view.

## 1. Introduction:

Ulam-Hyers stability studies the following question: Suppose one has a function y(t) which is close to solve an equation. Is there an exact solution x(t) of the equation which is close to y(t)?.Mathematically the following system can be studied ([3], [11]):

$$\frac{dx}{dt} = f(x) \qquad (1)$$

the system (1) is Ulam-Hyers (UH) stable if it has an exact solution and if $\forall \varepsilon > 0$ there is $\delta > 0$ such that if $x_a(t)$ is an approximation for the solution of (1) then there is an exact solution x(t) of (1) which is close to $x_a$ i.e.,

$$\left\| \frac{dx_0}{dt} - f(x_a(t)) \right\| < \delta \Rightarrow \|x(t) - x_0(t)\| < \varepsilon \, \forall t > 0. \qquad (2)$$

This definition has applicable significance since it means that if one is studying an UH stable system then one does not have to reach the exact solution (which usually is quite difficult or time consuming). All what is required is to get a function which satisfies (2). UH stability guarantees that there is a close exact solution. This is quite useful in many applications e.g. numerical analysis, optimization, biology and economics etc., where finding the exact solution is quite difficult. It also helps, if the stochastic effects are small, to use deterministic model to approximate a stochastic one.

We begin by realizing that UH stability is independent of the more familiar Lyapunov stability which states that the system (1) is Lyapunov stable if both x(t),y(t) are exact solutions of (1) and for all ε>0 there is δ>0 such that $|x(0)-y(0)|<\delta$ implies $|x(t)-y(t)|<\varepsilon$ for all t>0.

A known counter-example proving the independence of the two concepts is the system:

$$\frac{dx}{dt} = ax(t), \qquad a > 0 \ \cos s \tan t \qquad (3)$$

whose x=0 solution is Lyapunov unstable while it is UH stable [12].

UH stability has been studied for functional equations [16], and linear differential equations ([11], [17]). In sec.2 local UH stability for nonlinear differential and difference equations will be studied by proposing a concept analogous to local Lyapunov stability. Applications in the logistic (difference and differential) equation, SIS epidemic model, Cournot model in economics and a reaction-diffusion model are given in sec.3.

2.Local UH stability for nonlinear systems

Consider systems (1), (2), assume

$$y(t) = x(t) + h(t) \qquad (4)$$

also assume that h(t) is small hence linearize in it. Substituting in (1), (2) one finally gets

$$h(t) \le \delta f(x) \int \frac{dx}{f^2(x)}. \qquad (5)$$

Thus we have:

**Proposition (1):** The system (1) is locally UH stable if there is a constant K such that

$$\left| f(x) \int \frac{dx}{f^2(x)} \right| < K. \qquad (6)$$

Similarly consider the discrete system

$$x(t+1)=f(x(t)), \quad t=0,1,2,..., \quad |y(t+1)-f(y(t))|<\delta \qquad (7)$$

again assuming

$$y(t)=x(t)+g(t), \quad t=0,1,2,..., \qquad (8)$$

assume g(t) is small hence linearize in it one gets the following:

**Proposition (2):** A sufficient condition for the system (7) to be locally UH stable is that there is constant K such that

$$\left|\frac{df(x)}{dx}\right|<K<1. \qquad (9)$$

## 3  Applications:

1) Logistic differential equation:

$$\frac{dx}{dt}=rx(1-x), \qquad r>0 \text{ cons} \tan t. \qquad (10)$$

Applying proposition (1) the system (10) is locally UH stable if there is a constant K such that

$$|2x-1+2x(1-x)\log(|x/(x-1)|)|<K\cong 1.2, \forall x\in[0,1]. \qquad (11)$$

## 2) Logistic difference equation:

$$x(t+1)=rx(t)[1-x(t)], \quad r>0 \text{constant}, x\in[0,1]. \qquad \textbf{(12)}$$

Applying (9) the system (12) is locally UH stable if there is a constant K such that:

0<r<K<1.

3) SIS infection model in constant population:

The system is given by the equations:

$$\frac{dS}{dt}=-aSI ,$$
$$\frac{dI}{dt}=-aSI -bI, \qquad (13)$$

where a population of size N is divided into susceptible and infective (infectious) individuals, with numbers denoted by S(t),I(t), respectively. Thus N=S+I, a,b>0.

Rescaling by N one gets (10) hence the system (13) is locally UH stable. This is important since assessing the real number of infections can be extremely difficult. Also stochastic effects can be significant. Thus if the deterministic model is close enough to reality its convergence to an exact solution is guaranteed by the local UH stability.

**4) Wave solution of a reaction-diffusion system:**

The system is given by

$$\frac{\partial u}{\partial t} = \nabla^2(u) + f(u), \qquad (14)$$

where f(u) is differentiable function. Looking for a wave solution in one dimension

$$u=u(z), z=-ct+x,$$

hence one gets

$$((d^2u(z))/(dz^2))+c((du)/(dz))+f(u)=0. \qquad (15)$$

Using the procedure in proposition (1) one gets that (15) is locally UH stable if there is a constant K such that

$$((df)/(du))>K>0. \qquad (16)$$

5) A model for economic monopoly with constant output:

Monopoly [13] is the case where the market is controlled by one firm p. In such models one typically maximizes the profits. Consider a firm producing q(t) of a certain product at time t with profit function

$$P(q(t))=q(t)[a-cq(t)]$$

Using the standard approach to Cournot economic dynamical systems with bounded rationality

$$q(t+1)=q(t)+bq(t)((dP(q(t)))/(dq(t)))$$

(where bq(t) is a measure of the bounded rationality) one finally gets the coupled dynamical system

$$q(t+1)=q(t)[1-b(2cq(t)-a)] \qquad (17)$$

Rescaling one finds that the system is equivalent to (12). Hence the system (17) is locally UH stable.

Concluding Ulam-Hyers stability concept is quite significant in realistic problems in numerical analysis, biology and economics. A generalization to nonlinear systems is proposed and applied to the logistic equation (both differential and difference), SIS epidemic model Cournot model in economics and a reaction diffusion equation.

It is important to notice that there are many applications for UH stability in other topics e.g. in nonlinear analysis problems including differential equations and integral equations ([6], [7], [15]). The first time the "Hyers-Ulam-Rassias stability" was used in Partial Differential equations was in [14]. Hyers-Ulam-Rassias stability" was used in Ordinary Differential Equations in the papers ([9], [10]).

Further applications exist in the following references ([1], [2], [4], [5], [8]).

We thank Prof. Th. M. Rassias for his help.